\documentclass[twocolumn]{openjournal}

\usepackage{xcolor}
\usepackage[utf8]{inputenc}
\usepackage{hyperref}
\hypersetup{
    unicode,
    colorlinks=true,
    linkcolor=linkcolor,
    citecolor=linkcolor,
    filecolor=linkcolor,
    urlcolor=linkcolor,
}
\definecolor{linkcolor}{rgb}{0.0,0.3,0.5}
\DeclareGraphicsExtensions{.bmp,.png,.jpg,.pdf}
\usepackage{verbatim}

\newcommand{\photu}{photon units}

\newcommand{\galex}{{\it GALEX}}

\newcommand{\halpha}{H$\alpha$}
\newcommand{\nh}{{\it New Horizons}}
\newcommand{\chisq}{$\chi^{2}$}
\newcommand{\ebv}{E(B~-~V)}
 
\newcommand{\asec}{\hbox to 1pt{}\rlap{$^{\prime\prime}$}.\hbox to 2pt{}}
\newcommand{\amin}{\hbox to 1pt{}\rlap{$^{\prime}$}.\hbox to 1pt{}}
\newcommand{\adeg}{\hbox to 1pt{}\rlap{$^{\circ}$}.\hbox to 2pt{}}
\newcommand{\degree}{\hbox to 1pt{}\rlap{$^{\circ}$}\hbox to 2pt{}}

\providecommand{\apjl}{Astrophys.\ J.\ Lett.}

\providecommand{\apjs}{Astrophys.\ J.\ Suppl.}
\providecommand{\aap}{Astron.\ Astrophys.}

\begin{document}

\title{The Cosmic Ultraviolet Background at the Galactic Poles}

\author{Jayant Murthy}
\email{jmurthy@yahoo.com}
\affiliation{Indian Institute of Astrophysics, Bengaluru 560 034, India.}

\author{James Overduin}
\affiliation{Towson University, Towson, Maryland 21252, USA.}

\author{Raj Manas}
\author{Amit Pathak}
\affiliation{Department of Physics, Banaras Hindu University, Varanasi 221005, India.}

\author{Snehanshu Saha}
\affiliation{School of AI and CSE, Mahindra University, Hyderabad 500100, India (On lien from BITS Pilani K K Birla Goa Campus).}

\author{Richard C. Henry}
\affiliation{Johns Hopkins University, Dept. of Physics and Astronomy, Baltimore, MD 21218, USA.}

\begin{abstract}

We have used archival \galex\ data to separate the cosmic ultraviolet background at the Galactic Poles into two components: the dust scattered light and an offset. We have modeled the dust-scattered light using a single-scattering model finding $1 \sigma$ limits of 0.54 -- 0.71 for the albedo ($a$) and 0.74 -- 0.83 for the phase function asymmetry factor ($g$) at 1530 \AA\ and 0.66 -- 0.73 for $a$ and 0.71 -- 0.77 for $g$ at 2360 \AA, that is, the grains are moderately reflective and highly forward-scattering.

The offsets are 277 -- 284 \photu\ at 1530 \AA\ and 513 -- 520 \photu\ at 2360 \AA. After accounting for all known Galactic and extragalactic sources, we are left with an excess of $165 \pm 18$ \photu\ at 1530 \AA\ and $346 \pm 34$ at 2360 \AA. The offsets are constant over these regions with variances of 20 -- 30 \photu.
Both the intensity and shape (NUV-to-FUV ratio) of this unexplained signal appear consistent with predictions based on emerging dark-matter candidates known as Axion Quark Nuggets (AQN).

\end{abstract}


\maketitle

\section{Introduction}\label{sec:intro}

The cosmic ultraviolet background (CUVB) is a mix of Galactic and extragalactic sources \citep{Murthyreview2009}, with the primary component at low Galactic latitudes being dust-scattered starlight \citep{Murthy2010}. Other contributors to the diffuse Galactic light (DGL) include line emission from hot gas at high Galactic latitudes \citep{Feldman_hotgas1981, Martin1990_lines}, and fluorescent emission from molecular hydrogen in the Lyman and Werner bands \citep{Martin1990_h2}, if the hydrogen column density is greater than about $5 \times 10^{20}$ cm$^{-2}$ \citep{Savage1977}. The primary contributor to the extragalactic background light (EBL) is the redshifted emission of galaxies \citep{Driver2016}.

Observations of the CUVB began in the late 1960s with the rocket observations of \citet{Hayakawa1969} and have continued to the present, many with sounding rockets but more recently from spacecraft. We have listed all published observations of the CUVB, including those studies that used archival data, in Appendix \ref{app:prevobs} (Table \ref{tab:agval}). It was recognized early that the best locations to untangle the different components would be near the Galactic Poles, where the dust-scattered radiation is minimized, and we have noted those observations in Table \ref{tab:agval}.

The earliest observations found a surface brightness of 200 - 300 photons cm$^{-2}$ s$^{-1}$ sr$^{-1}$ \AA$^{-1}$ (hereafter \photu) at the North Galactic Pole and postulated this to be the EBL, given that the Galactic emission was thought to be close to zero. More recent investigations using data from the far ultraviolet (FUV: 1350 -- 1800 \AA) and the near ultraviolet (NUV: 1800 -- 2800 \AA) imagers on the {\it Galaxy Evolution Explorer} satellite (\galex) \citep{Hamden2013, Akshaya2018, Akshaya2019} have shown that the CUVB may be deconvolved into two components:
\begin{itemize}
    \item Dust-scattered light linearly correlated with the reddening (\ebv).
    \item An offset of 200 -- 300 \photu\ in the FUV and 500 -- 650 \photu\ in the NUV.
\end{itemize}
About 25 -- 30\% of the offset may be assigned to the EBL \citep{Driver2016} and another 15\% to emission from hot gas and two-photon emission (see Section \ref{sec:contributors}). The remainder, about 60\% of the total observed CUVB in the FUV and 70\% in the NUV, is still of unknown origin, although there have been recent suggestions that it is a signature of axion quark nuggets \citep{Zhitnitsky2022}.

In this work, we use archival \galex\ data at the Galactic Poles ($|b| > 80\degree$) to better constrain the dust-scattering and the offsets. We have used a single-scattering model of dust-scattering to show that the grains have an albedo ($a$) of 0.54 -- 0.71 at the $1\sigma$ level and a phase function asymmetry factor ($g$) of 0.74 -- 0.83 in the FUV and 0.66 -- 0.73 and 0.71 -- 0.77, respectively, in the NUV; that is, they are strongly forward-scattering with a high reflectivity. We confirm the offsets with  strong $1\sigma$ limits of 277 -- 284 \photu\ in the FUV and 513 -- 520 \photu\ in the NUV, and find that they do not vary over the observed field. There are small differences between the offset at the NGP and the SGP, but these are likely due to the difficulties of the measurement, rather than real astrophysical differences.

\section{Data}

\begin{figure*}
    \includegraphics[width=0.333\textwidth]{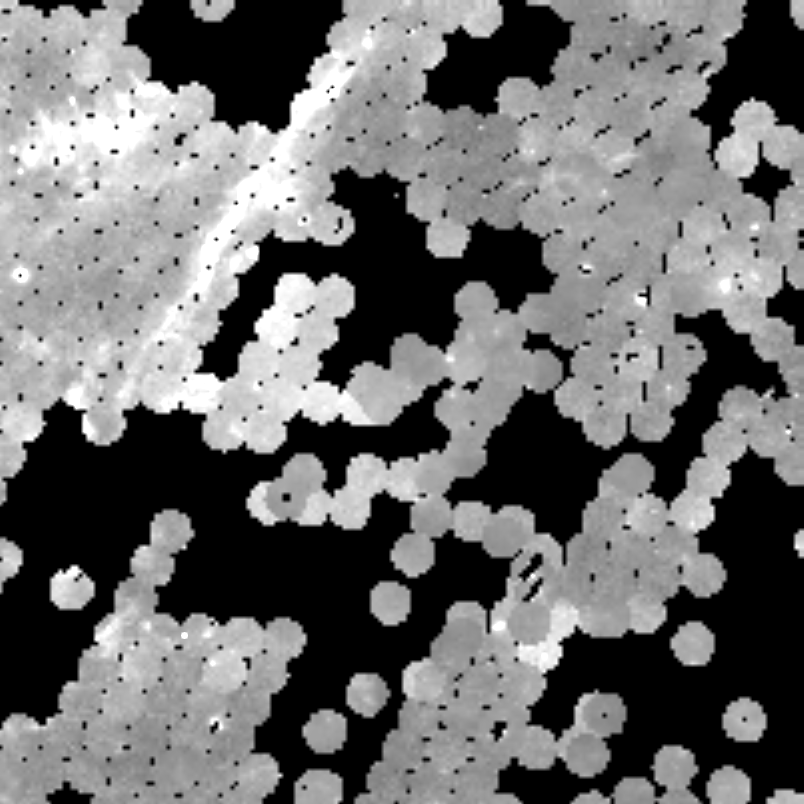}
    \includegraphics[width=0.333\textwidth]{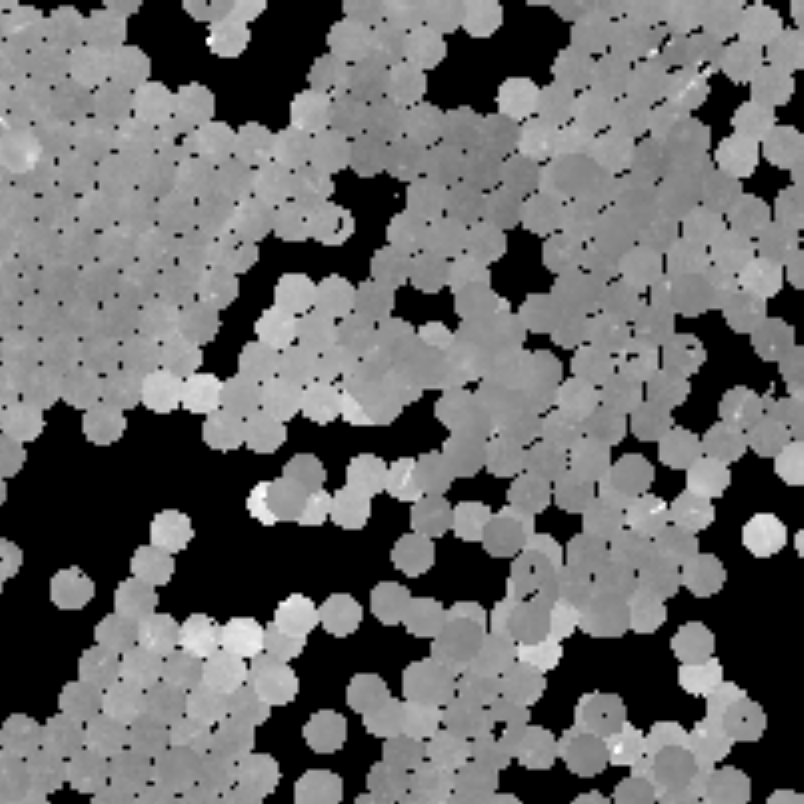}
    \includegraphics[width=0.333\textwidth]{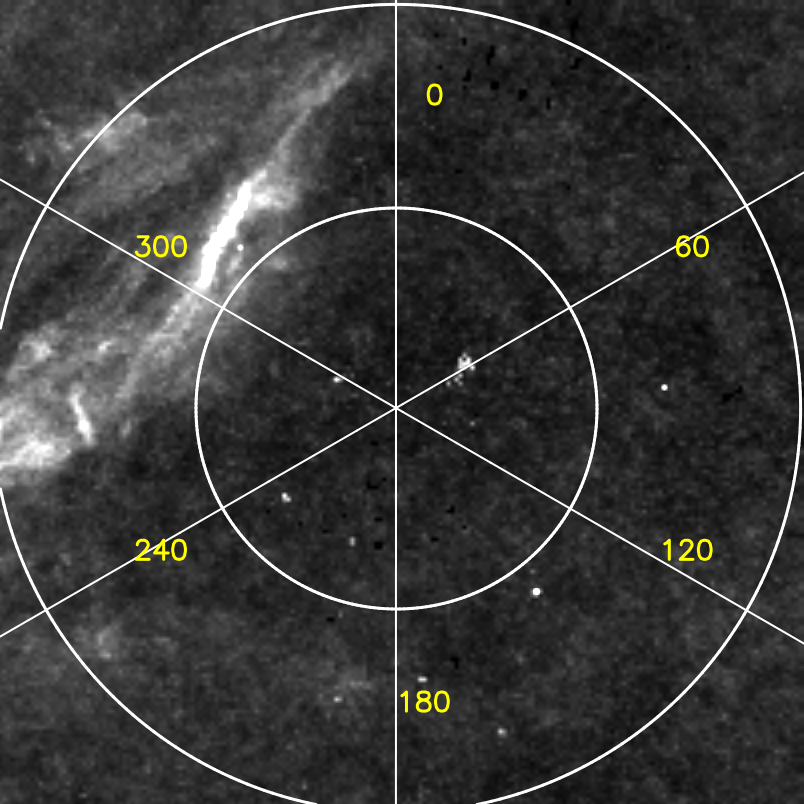}
    \includegraphics[width=0.333\textwidth]{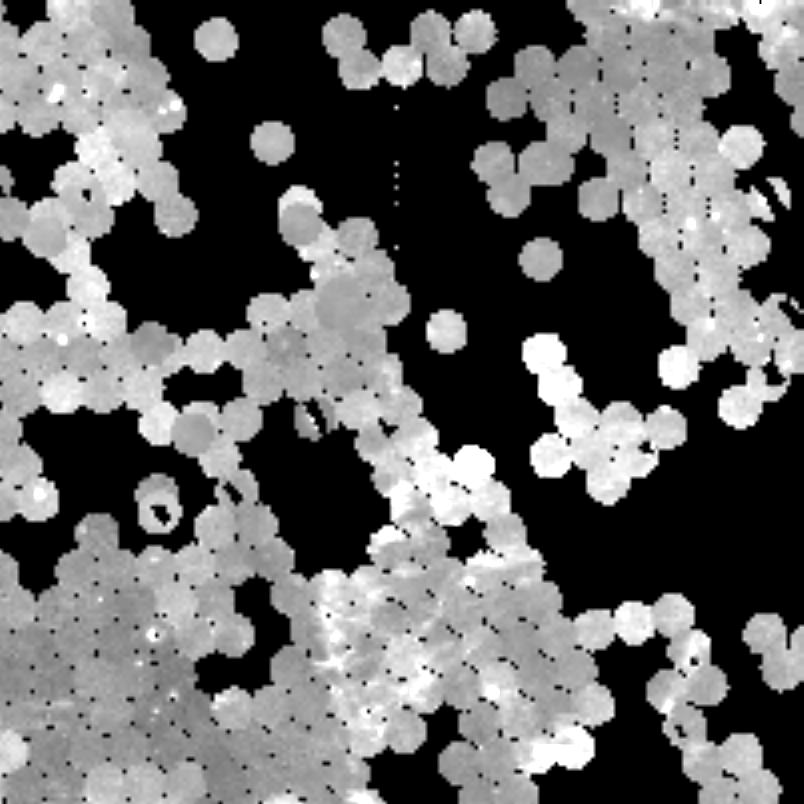}
    \includegraphics[width=0.333\textwidth]{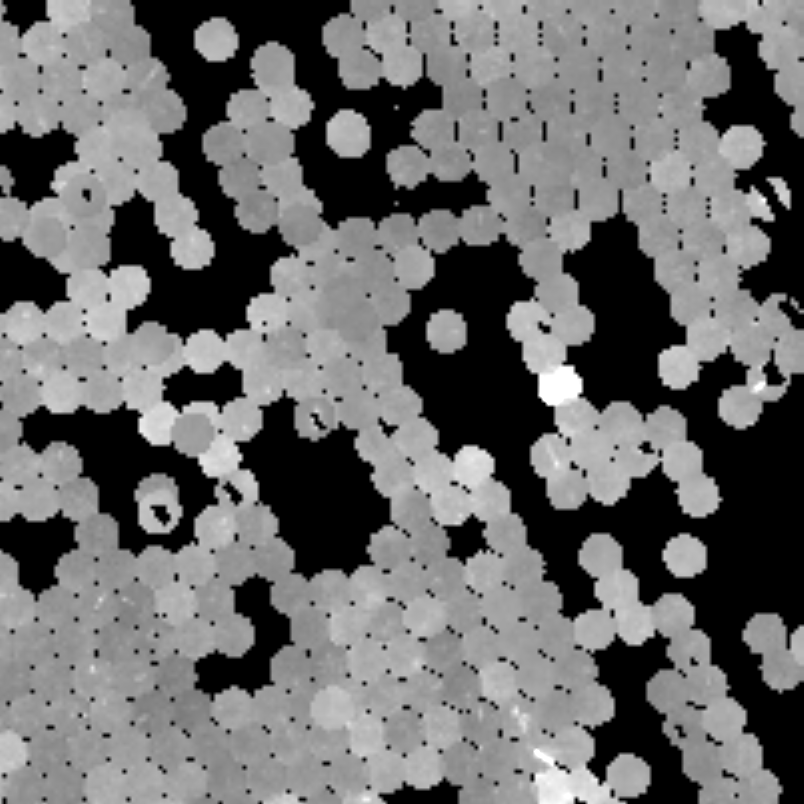}
    \includegraphics[width=0.333\textwidth]{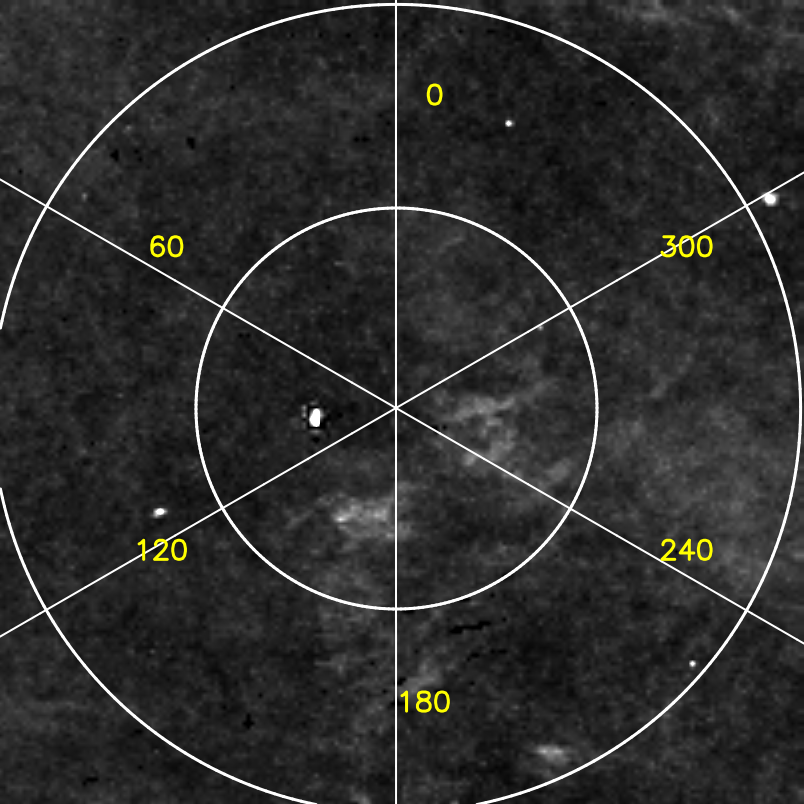}
    \caption{FUV (left), NUV (centre), and Planck \ebv\ (right) for the NGP (top) and SGP (bottom). The coordinates are shown in the \ebv\ maps with the pole at the center and the 80\degree\ and 85\degree\ lines of latitude plotted. Black patches in the UV images are where no observations were taken, often because of nearby bright stars.}
    \label{fig:polar_maps}
\end{figure*}

\citet{Murthy2014apj} created maps of the CUVB from \galex\ observations in the FUV and NUV. He subtracted the foreground emission, masked out point sources, and binned the data into $2'$ bins. These data are available from the Mikulski Archive for Space Telescopes site at \url{https://archive.stsci.edu/prepds/uv-bkgd/}, with higher spatial resolution ($15"$) data available from Zenodo (\url{https://doi.org/10.5281/zenodo.13337911}).
The reddening data come from \citet{PlanckDust2016} with a spatial resolution of about $5'$. We have rebinned the data into $6'$ bins and imaged them in Fig. \ref{fig:polar_maps}.

\begin{figure*}
    \includegraphics[width=0.5\textwidth]{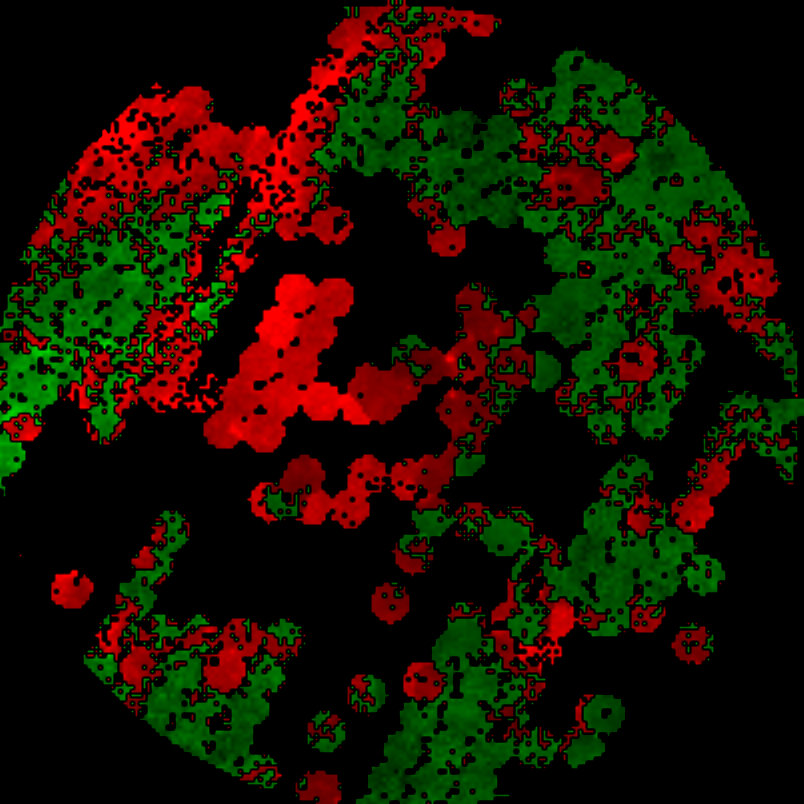}
    \includegraphics[width=0.5\textwidth]{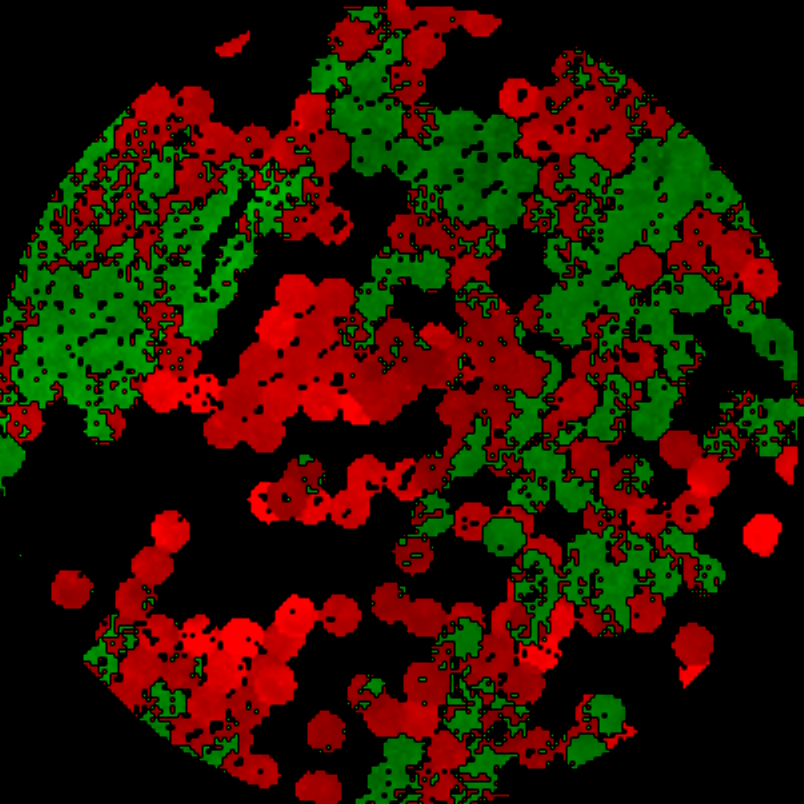}
    \includegraphics[width=0.5\textwidth]{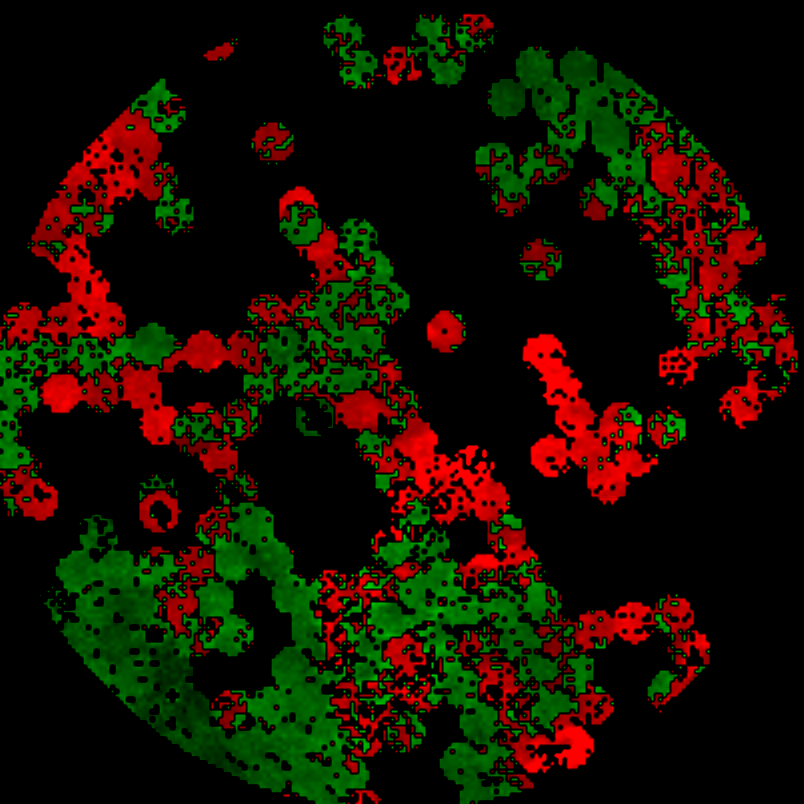}
    \includegraphics[width=0.5\textwidth]{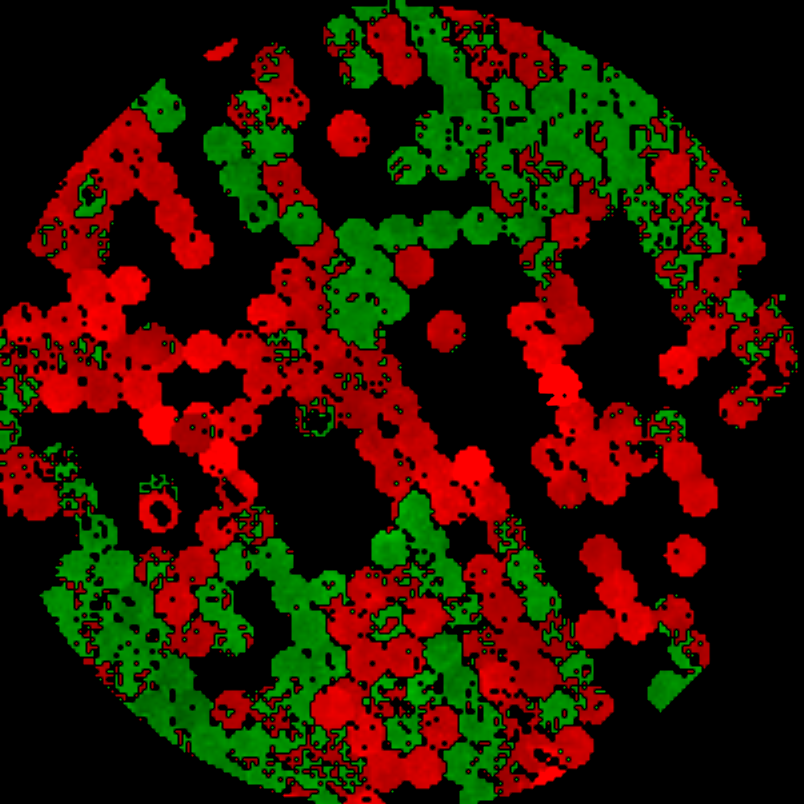}
    \caption{The NGP (top) and SGP (bottom) in the FUV (left), and NUV (right) where the black areas either contained no data or had an \ebv\ greater than 0.1 mag; the red areas show an excess of emission over the linear fit; and the green areas were used in this work. 
    }
    \label{fig:polar_mask}
\end{figure*}

\begin{figure*}
    \includegraphics[width=\textwidth]{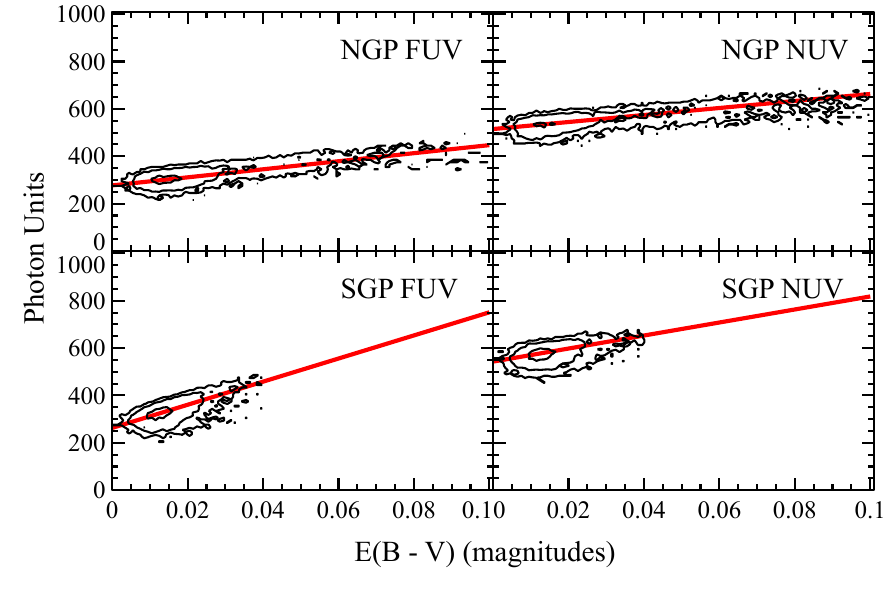}
    \caption{Density plots for \galex\ UV radiation plotted against \ebv, where we have rejected those points that show additional contributors. The contours represent bins with the outermost contours set for bins with .01\%, and .1\%. We have also plotted the innermost contours at levels of 1\%, and 10\% of the total number of points but they are not visible in most of the plots. The best-fit lines from Eq. \ref{eq:CUVB_relation} are shown in each panel.}
    \label{fig:uv_ebv_corr}
\end{figure*}

\begin{table}
\centering
\caption{UV-EBV Fits}
\label{tab:corr}
\begin{tabular}{llll}
\hline
Band & p$^{a}$ & A$^{b}$ & O$^{c}$\\
\hline
NGP FUV & 0.721 & $ 1489 \pm    17 $ & $  270 \pm 0.32 $\\
NGP NUV & 0.701 & $ 1461 \pm    12 $ & $  503 \pm 0.25 $\\
SGP FUV & 0.570 & $ 3786 \pm    46 $ & $  260 \pm 0.69 $\\
SGP NUV & 0.424 & $ 2666 \pm    36 $ & $  523 \pm 0.58 $\\
\hline
\multicolumn{4}{l}{$^{a}$Correlation coefficient.}\\
\multicolumn{4}{l}{$^{b}$\photu\ mag$^{-1}$ .}\\
\multicolumn{4}{l}{$^{c}$\photu.}\\
\end{tabular}
\end{table}

The dust at the Galactic Poles has a low column density and we have approximated the CUVB with a linear term proportional to the reddening and a constant offset:
\begin{equation}
    CUVB = A \times E(B - V) + O
    \label{eq:CUVB_relation}
\end{equation}
where $A$ is the slope in units of \photu\ mag$^{-1}$ and $O$ is the offset in \photu. This has been the usual practice in the literature and the values listed in Table \ref{tab:agval} depend on this assumption, with the sole exception of \citet{Murthy2025}. 

We have found that the offset is not constant over the entire field because of other Galactic contributors, notably molecular hydrogen fluorescence or nearby stars and, therefore used a two-step process in which we fit the data with Eq. \ref{eq:CUVB_relation} and then rejected those points more than $1\sigma$ greater than the linear fit. The rejected points are shown in red in Fig. \ref{fig:polar_mask} and will be discussed in Section \ref{sec:exc_reg}.

\begin{table}
\centering
\caption{Largest stellar contributors to CUVB ($a = 0.6, g=0.8) $.}
\label{tab:ngp_stars}
\begin{tabular}{lllrrll}
\hline
HIP$^{a}$ & Star & Sp. Type$^{a}$ & $l$ ($^{\circ}$) & $b$ ($^{\circ}$) & D$^{ab}$ & \%$^{c}$\\
\hline
\multicolumn{7}{c}{NGP}\\
\hline
65474 & $\alpha$ Vir       & B1 III-IV            & 316.1 & $+50.8$ & 77 & 8.3\\
51624 & $\rho$ Leo              & B1 Iab                      & 234.9 & $+52.8$ & 900 & 3.9\\
60718 & $\alpha$ Cru       & B0.5 IV     & 300.1 & $-0.4$  & 106 & 2.5\\
78401 & $\delta$ Sco    & B0.3 IV             & 350.1 & $+22.5$ & 136 & 2.4\\
26727 & $\zeta$ Ori      & O9.5 Iab      & 206.5 & $-16.6$ & 387 & 2.4\\
\hline
\multicolumn{7}{c}{SGP}\\
\hline
4577  & $\alpha$ Scu         & B7 IIIp            & 268.1 & $-87.3$ & 216 & 16\\
183   & $\zeta$ Scu          & B5 V               & 16.6  & $-78.9$ & 154 & 3.8\\
27366 & $\kappa$ Ori        & B0.5 Ia            & 214.5 & $-18.5$ & 198 & 3.0\\
7588  & $\alpha$ Eri     & B3 Vpe   & 290.8 & $-58.8$ & 43  & 1.9\\
4427  & $\gamma$ Cas        & B0.5 IVe           & 123.6 & $-2.2$  & 168 & 1.8\\
\hline
\multicolumn{7}{l}{$^{a}$ From \citet{Perryman1997}.}\\
\multicolumn{7}{l}{$^{b}$ Distance in pc.}\\
\multicolumn{7}{l}{$^{c}$ Percentage of total dust-scattered light.}\\

\end{tabular}
\end{table}

The remaining points (shown in green in Fig. \ref{fig:polar_mask}) are those where there is a good linear correlation between the \ebv\ and the UV surface brightness (Fig. \ref{fig:uv_ebv_corr}), with the correlations and fitted parameters listed in Table \ref{tab:corr}. Although we have listed formal $1 \sigma$ uncertainties for the slope and the offset in the Table, these are nonphysical and are underestimated because the individual data points are not independent (discussed below). The small difference in the offsets between poles is likely representative of the actual uncertainty. These results are also consistent, again given the uncertainty, with earlier results (Table \ref{tab:agval}). 

The interstellar radiation field (ISRF) is about twice as strong in the SGP and hence the slopes of the UV/\ebv\ lines are a factor of about 2 greater in the South. We have tabulated the brightest stars in Table \ref{tab:ngp_stars}, finding that the brightest 5 stars contribute about 20\% of the scattered photons in the NGP and 27\% in the SGP, with $\alpha$ Scu being very near the SGP.

\subsection{Modeling the dust-scattered light}

We have modeled the dust-scattered light using the single-scattering formulation of \citet{Murthy2025}. Stars with their positions, magnitudes, and spectral types were taken from the Hipparcos catalog \citep{Perryman1997} and the spectrum of each star was modeled using TLUSTY spectral profiles \citep{Lanz2003, Lanz2007}. The amount of dust between the star and the location of the dust was calculated using the 3-dimensional dust map of \citet{Green2019} and the starlight was extincted with the curve of \citet{Draine_scat2003}. Finally, we convolved the ISRF with the dust distribution along the line of sight, assuming that the column density from the Planck \ebv\ \citep{PlanckDust2016} was distributed along a column, with the density falling exponentially with a scale height of 125 pc \citep{Marshall2006} and a cavity of radius 50 pc around the Sun \citep{Welsh2010}. As \cite{Murthy_dustmodel2016} showed, the observed surface brightness is an integral over the entire column and is not sensitive to the details of the dust distribution.

We assumed that the scattering function was given by \citet{Henyey1941} with the optical constants as free parameters: the albedo ($a$) and the phase function asymmetry factor ($g$), and calculated the expected scattered light as a function of the optical constants. The offset, extincted by the total amount of dust, is added to the dust-scattered light:
\begin{equation}
        CUVB = D(\lambda, \ebv, a, g, gl, gb) + O \exp(-\tau)
    \label{eq:dust_scatter}
\end{equation}
where $D$ is the dust-scattered light from the model, $O$ is the offset, and $\tau$ is the optical depth along the line of sight. Note that the dust-scattered light ($D$) is a function of the total reddening along the line of sight and varies across the sky, depending on the local ISRF.

\subsection{Fitting the data}

We tried using the \citet{Lampton1976} formulation to derive limits on the parameters but found that it gave unrealistically low uncertainties, because the effective spatial resolution of the dust-scattered light is much coarser than the $6'$ bin size of the data. The incident interstellar radiation field is smeared by scattering in the dust column along the line of sight with an autocorrelation analysis suggesting an appropriate spatial resolution of about $1 \degree$. The effect on our data analysis was that the data points were not independent and, hence, were not amenable to a simple \chisq\ analysis.

A robust method of estimating uncertainties is the block bootstrap (jackknife) method introduced by \citet{Kuensch1989} and \citet{Liu1992}, with astronomical applications discussed by \citet{hynes2016, Thomsen2022} and \citet{Saydjari2026}. This method involves organizing the data into blocks of data and then replacing each block with another block at random, allowing duplications and deletions. In our case, we defined blocks of different spatial dimensions, where each block consisted of the \ebv, the data (FUV or NUV) and the model ($D[\lambda, \ebv, a, g, gl, gb]$ from Eq. \ref{eq:dust_scatter}). The block size should be chosen large enough to encompass the spatial correlation but not so large that there are too few blocks to yield representative distributions. We further allowed for the measurement uncertainties by picking the \ebv\ and the UV values randomly from a normal distribution with a width given by the uncertainty for each pixel.
We fit the resulting distribution with our model (Eq. \ref{eq:dust_scatter}), 
finding the best-fit optical constants ($a$ and $g$) and offset ($O$) for each run.

\begin{figure}
    \includegraphics[width=1\columnwidth]{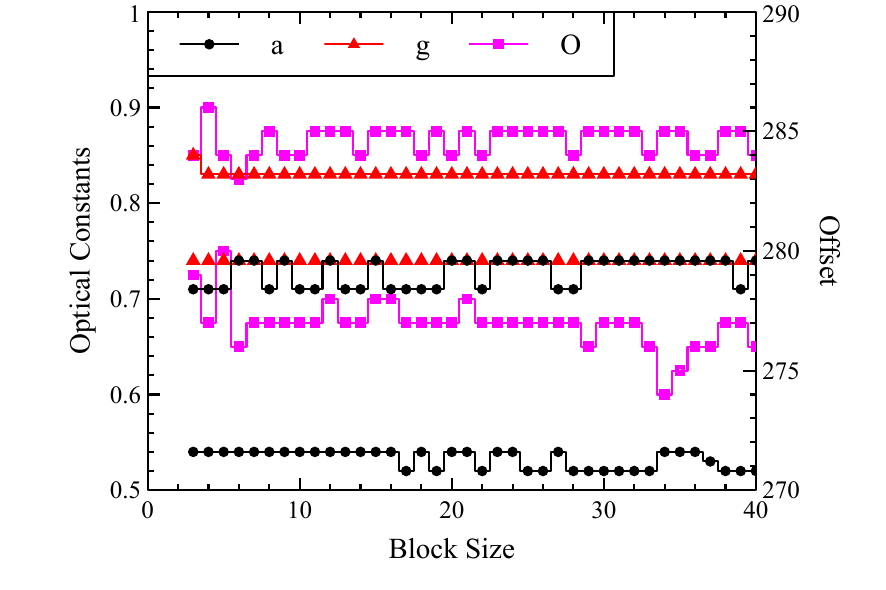}
    \caption{$1 \sigma$ variance in the optical constants (left axis) and the offsets (right axis in \photu) for the NGP in the FUV band as a function of the block size in pixels (1 pixel is 6 arcminuts), where the key is given in the  Figure.}
    \label{fig:block_size}
\end{figure}

 We repeated this process over several thousand runs, selecting the blocks randomly for each iteration, with the range in the best-fit parameters across all the runs defining the uncertainties, i.e., 68.3\%, 95.4\%, and 99.7\%\ of the values fell within the allowed ranges for $1\sigma, 2\sigma,$ and $3\sigma$ uncertainties, respectively. We have plotted the size of the allowed $1 \sigma$ regions for the optical constants ($a$ and $g$) and the offsets ($O$) in Fig. \ref{fig:block_size} for a range of block sizes. If we choose the block size too small, we will not sample independent regions and the allowed uncertainties will be underestimated; if we choose the block size too large, the distribution of pixels will not represent the actual distribution and the uncertainties will be too large. We have chosen a block size of 18 pixels, corresponding to 108 arcmin, but note that the results are not particularly dependent on the block size from roughly 60 -- 150 arcmin.

\section{Results}
\subsection{Optical Constants}
\begin{figure}
    \includegraphics[width=1\columnwidth]{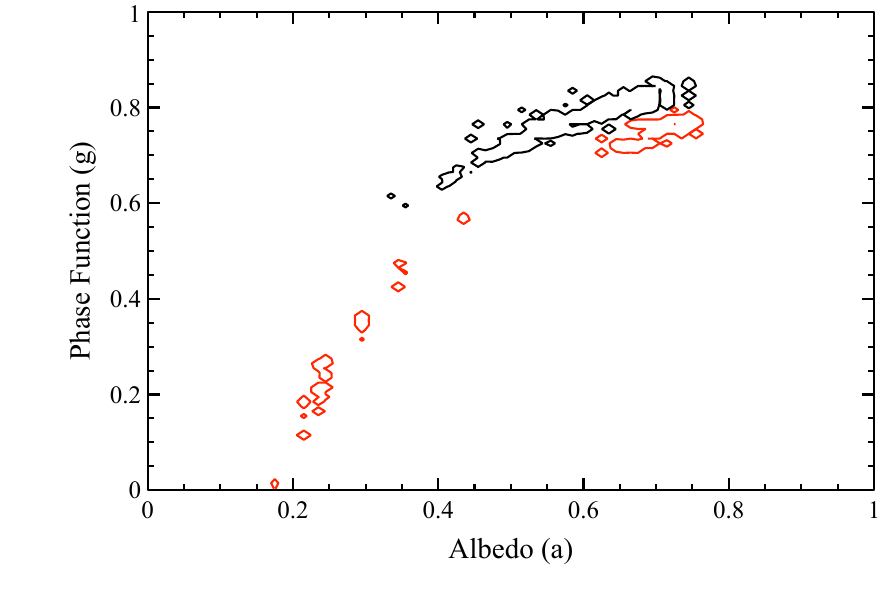}
    \caption{$3 \sigma$ contours of the optical constants in the FUV (black contours) and NUV (red contours). We have only shown the contours for the NGP where the broader range of \ebv\ allowed us to set tighter limits on the optical parameters. The allowed range is larger but consistent when the SGP is considered.}
    \label{fig:block_ag}
\end{figure}

\begin{table}[htbp]
\centering
\caption{Fitted Parameters}
\label{tab:params}
\begin{tabular}{llll}
\hline
Band & $a$ & $g$ & O$^{a}$\\
\hline
NGP FUV & 0.54 -- 0.71 & 0.74 -- 0.83 & 277 -- 284\\
SGP FUV & 0.24 -- 0.68 & 0.47 -- 0.88 & 280 -- 293\\
NGP NUV & 0.66 -- 0.73 & 0.71 -- 0.77 & 513 -- 520\\
SGP NUV & 0.19 -- 0.74 & 0.13 -- 0.83 & 533 -- 554\\
\hline
\multicolumn{4}{l}{$^{a}$\photu.}\\
\multicolumn{4}{l}{$^{b}$ One point at $a = 0.2$ and $g = 0.2$ has been excluded.}\\
\multicolumn{4}{l}{Note: Allowed range, independent of other parameters.}\\

\end{tabular}
\end{table}

We have plotted the allowed $3 \sigma$ range in the optical parameters for the FUV and the NUV in the NGP in Fig. \ref{fig:block_ag}, noting that the limits in the SGP are consistent but broader because of the smaller range in the \ebv\ values. Although $a$ and $g$ are strongly correlated, we have tabulated their independently determined $1 \sigma$ limits in Table \ref{tab:params}. We find moderately reflective, highly forward-scattering grains, consistent with the range derived by \citet{Murthy2025}. 

We have tabulated earlier results for the optical constants in Appendix \ref{app:prevobs}, noting that the range in the derived values reflects the uncertainty in the data and, more importantly, the models. As \citet{Mathis2002} have noted, observations of both the diffuse Galactic light and of reflection nebulae suffer from uncertain geometries of the stars and the scattering grains. 

\subsection{Offsets}

\subsubsection{Observations}

The offset is that part of the CUVB that is not correlated with the reddening and, as such, its derivation is robust and does not depend on the model for interstellar scattering. We have tabulated the $1 \sigma$ range for the offsets in Table \ref{tab:params}, finding limits very similar to previous determinations (Table \ref{tab:agval}). The range in the SGP is slightly different but, we believe, consistent with that found in the NGP given the smaller range in \ebv\ sampled in the SGP. We find variances of 20 -- 30 \photu\ over both Poles, similar to the cited uncertainties in the \galex\ data \citep{Murthy2014apj}, and thus find no indication of any spatial structure in the offsets.

\subsubsection{Contributors}\label{sec:contributors}
\begin{itemize}
    \item The EBL is from the integrated light of external galaxies and is estimated to be $73 \pm 16$ and $158 \pm 34$ \photu\ in the FUV and NUV, respectively \citep{Driver2016}.
    \item Resolved stars have already been removed from the data \citep{Murthy2014apj}. The only stars that can contribute to either \galex\ band are hot stars (O, B, and A stars) and white dwarfs hotter than about 12{,}000 K. These stars would be detected as point sources out to distances of several kiloparsecs, well outside the Galaxy at these latitudes, and would have been flagged and excluded from our diffuse data files. Thus we expect no contribution from unresolved stars in the \galex\ bands.
    \item The brightest emission line in the FUV is the C~IV doublet (1548/1550 \AA) which contributes $20 \pm 5$ \photu\ to the diffuse flux, based on observations of 4000 - 6000 ph cm$^{-2}$ s$^{-1}$\citep{Martin1990_lines, Korpela2006, Jo2019}. There are no strong observational constraints on line emission in the NUV and there are no significant emission lines in this band from hot gas, because the emission lines of the common ions in the halo (C~IV, N~V, O~VI, Si~IV) are all at shorter wavelengths.
    \item Two-photon emission from hydrogen \citep{Deharveng1982, Martin1991_ebl, Reynolds1992} will contribute to the CUVB at a level proportional to the H$\alpha$ emission. The average \halpha\ emission over the NGP is 0.45 Rayleighs of which about 20\% is due to diffuse radiation scattered by interstellar dust \citep{Wood1999, Dong2011, Brandt2012}. We have used the formulation of \citet{Bracco2020} to calculate a contribution from two-photon emission of $22 \pm 5$ and $12 \pm 3$ \photu\ in the FUV and NUV, respectively. \citet{Kulkarni2022} predicted that there would also be a contribution  from geocoronal and interplanetary two-photon emission;  however, \citet{Murthy2025_alice} showed from \nh\ Alice observations that there was, in fact, no such emission.
    \item The source of the remaining $165 \pm 18$ \photu\ in the FUV and $346 \pm 34$ \photu\ in the NUV is still unknown, with a recent suggestion that it is axion quark nuggets (AQN) (Section \ref{sec:aqn_sec}).
\end{itemize}

\begin{table}[t]
\caption{Components of Offsets
\label{tab:summary}}
\begin{tabular}{lll}
\hline\hline
Source & Offset$^{a}$ & Ref.\\
\hline
\multicolumn{3}{c}{1530~\AA}\\
\hline
Observed  & $ 280 \pm 4$ & This work\\
EBL & $73 \pm 16$ & \citet{Driver2016}\\
CIV & $20 \pm 5$ & \citet{Martin1990_lines}\\
Two-photon & $22 \pm 5$ & This work\\
\hline
Excess & $165 \pm 18$\\
\hline
\multicolumn{3}{c}{2360~\AA}\\
\hline
Observed  & $ 516 \pm 4$ & This work\\
EBL & $158 \pm 34$ & \citet{Driver2016}\\
Two-photon & $12 \pm 3$ & This work\\
\hline
Excess & $346 \pm 34$\\
\hline
\multicolumn{3}{l}{$^a$ \photu.}\\
\end{tabular}
\end{table}

\subsection{Excluded Regions \label{sec:exc_reg}}

As discussed above, we have selected only those regions where the offset was constant (green areas in Fig. \ref{fig:polar_mask}). The red areas in the Figure show excess emission, each area of which may have different sources in the FUV and the NUV. For instance, Markkanen's Cloud \citep{Markkanen1979} may be dense enough to form molecular hydrogen which would radiate in the \galex\ FUV in the Werner bands but not in the NUV. Other areas may be due to nearby bright stars whose radiation would  light up the sky around the star. We plan to study these individually in a future work.

\section{NUV/FUV Ratio and a Possible Test of Axion Quark Nuggets as Dark-Matter Candidates }\label{sec:aqn_sec}

The UV excess reported here has been seen since the early days of rocket observations but its origin has remained mysterious \citep{Henry2015,Akshaya2018,Akshaya2019}. Recently, a plausible explanation has been identified in the form of Axion Quark Nuggets or AQNs \citep{Zhitnitsky2022}.
These are dark matter candidates whose existence would also solve two other outstanding mysteries in cosmology: baryon asymmetry (what happened to the antimatter?) and the ``coincidence problem'' (why are the densities of matter and dark matter comparable?).

Antimatter AQNs are composite particles whose contributions to the background radiation are broadband in nature, unlike the monochromatic spectra associated with the decay or annihilation of elementary particle dark-matter candidates such as massive neutrinos, axions, and supersymmetric WIMPs \citep{Overduin2008}. 
The mechanism for this radiation is thermal Bremsstrahlung or braking radiation from positrons in the ``electrosphere'' surrounding each antimatter nugget, which is heated to temperature $T$ by annihilations of antiquarks inside the nugget with baryons in the interstellar medium.
Both the peak wavelength and overall luminosity of this spectrum depend strongly on AQN mass. Heavier nuggets produce a hotter background but reduce its overall luminosity (since they imply a lower number density, given that the total mass density of dark matter is fixed). Lighter nuggets, by contrast, produce a background that is cooler but brighter.
Numerical simulations suggest the existence of a ``sweet spot'' for AQNs of mass near $300 \pm 200$~g whose spectrum would peak in the FUV with an intensity of 100-200 \photu\ (Fig. 8 of \citealt{Sekatchev2026}), similar to what we observe.

\begin{figure}
    \includegraphics[width=\columnwidth]{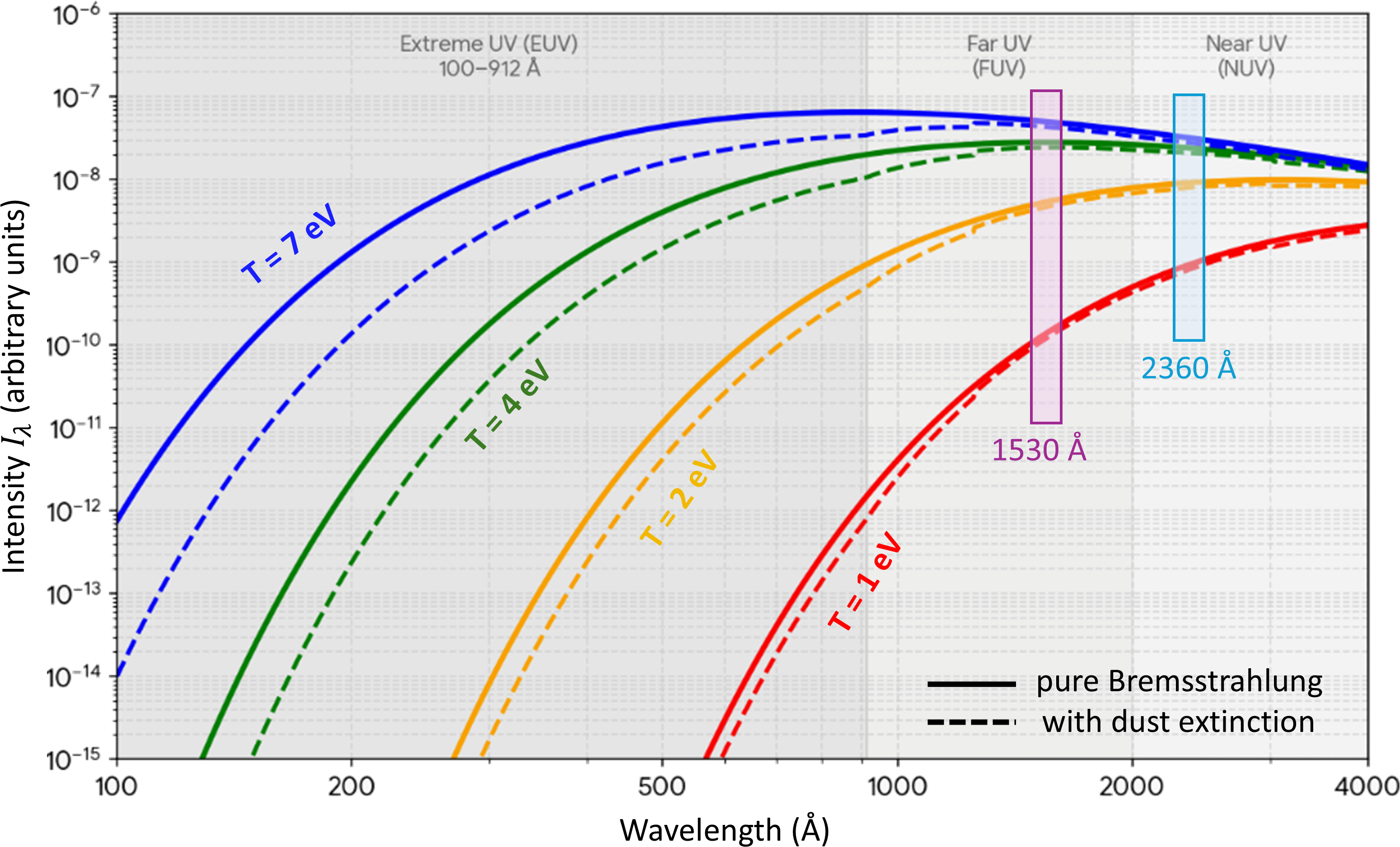}
    \caption{Spectra of the predicted signal from Bremsstrahlung radiation due to interactions of baryons in the ISM with dark-matter AQNs (Axion Quark Nuggets) surrounded by electrospheres of temperature $T$ (solid lines). Dashed lines suggest the effects of extinction by dust in the ISM. The ratio of NUV to FUV intensity is of particular interest (shaded rectangles).}
    \label{fig:aqns}
\end{figure}

A possible test of this hypothesis lies in the {\em ratio} of the intensities in the NUV and FUV bands.
Our unexplained excess of $165 \pm 18$ \photu\ at 1530 \AA\ (FUV) and $346 \pm 34$ \photu\ at 2360 \AA\ (NUV) implies a ratio of NUV/FUV$=2.1 \pm 0.5$.
The predicted intensity of free-free Bremsstrahlung radiation from a hot plasma at temperature $T$ can be expressed as a function of wavelength (i.e., in \photu) as
\begin{equation}
    I_{\lambda}\propto\frac{g_{FF}(\lambda,T)}{\lambda^2\sqrt{T}}\exp\left(-\frac{hc}{k_BT\lambda}\right) \;\;\; ,
\end{equation}
where the Gaunt factor $g_{FF}\approx$ const over the range considered here.
We plot this spectrum in arbitrary units in Fig.~\ref{fig:aqns} for $T=1,2,4$ and 7~eV (solid lines).
Also shown are simple estimates of the effects of interstellar dust extinction on the same four spectra following the widely-used Cardelli, Clayton and Mathis (CCM89) extinction law \citep{Cardelli1989}
with the standard extinction ratio $R_V=3.1$ and a low baseline visual extinction $A_V=0.1$~mag chosen to reflect low column densities along the line of sight to the Galactic poles (dashed lines).
Our FUV and NUV windows are indicated by shaded rectangles.
Extinction reduces overall intensity but does not appreciably change the NUV/FUV {\em ratios}, making this a potentially robust test of the AQN hypothesis.
These ratios do however depend sensitively on temperature. Based on the simplifying assumptions adopted here, observations of the diffuse galactic FUV background appear to favor models with $T\approx 2-3$~eV, consistent with theoretical expectations of $T\sim 1$~eV \citep{Sekatchev2026}. This finding could perhaps be used to constrain the AQN hypothesis.

\section{Conclusions}

We have studied the diffuse UV radiation at the Galactic Poles using a single-scattering model for the dust-scattered radiation. We find that the dust grains are reflective and highly forward-scattering with $1 \sigma$ limits of 0.54 -- 0.71 and 0.74 -- 0.83 for $a$ and $g$, respectively, in the FUV. The derived optical constants are similar in the NUV with $1\sigma$ limits on $a$ and $g$ being 0.66 -- 0.73 and 0.71 -- 0.77, respectively. It has been hard to derive the optical constants \citep{Mathis2002} and there are no consistent constraints on the optical constants of the dust grains in the UV.

The offsets are much better determined and we find $1 \sigma$ limits of 277 -- 284 and 513 -- 520 \photu\ in the FUV and NUV, respectively. There is no evidence for any variation in the offset between the Poles and we find that the variance is between 20 and 30 \photu\ over both Poles and both bands, similar to the expected variance from \galex\ data \citep{Murthy2014apj}. After subtraction of the expected contributors from Galactic and extragalactic sources, we find an excess of $165 \pm 18$ \photu\ in the FUV and $346 \pm 34$ \photu\ in the NUV.

These intensities appear consistent with the diffuse galactic background spectra produced by Bremsstrahlung radiation from interactions between baryons in the ISM and antimatter AQNs with masses between 100-500~grams. Moreover, the observed ratio of NUV to FUV intensities appears to favor models in which these interactions heat the positron electrospheres around these AQNs to temperatures between 2-3~eV, consistent with theoretical expectations.

We have identified a number of regions at either Pole with a much larger offset indicating local sources of the diffuse radiation, including molecular hydrogen fluorescence from Markkanen's Cloud \citep{Markkanen1979} and scattering from nearby stars such as $\alpha$ Scu, near the SGP. We will study these in a future work.

\section*{Acknowledgements}
Some of the data presented in this paper were obtained from the Mikulski Archive for Space Telescopes (MAST). STScI is operated by the Association of Universities for Research in Astronomy, Inc., under NASA contract NAS5-26555. Support for MAST for non-HST data is provided by the NASA Office of Space Science via grant NNX13AC07G and by other grants and contracts. This research has made use of the SIMBAD database, CDS, Strasbourg Astronomical Observatory, France. We have used the Gnu Data Language for the data analysis \citep{GDL2010, GDL2022}.

We have used Claude Opus for parts of the analysis. Claude suggested the use of the jackknife derivation of errors. In all cases, we read the original papers in wrote our own analysis programs. None of the text was written by an AI.

\clearpage
\begin{appendix}
\section{Previous Observations}\label{app:prevobs}
\begin{longtable}{llllll}
\caption{Previous Observations of the CUVB} \label{tab:agval} \\
\hline
Wavelength (\AA) & $a$ & $g$ & Offset$^{a}$ & Instrument & Reference\\
\hline
\endfirsthead

\multicolumn{6}{c}{{\tablename\ \thetable{} -- continued from previous page}} \\
\hline
Wavelength (\AA) & $a$ & $g$ & Offset$^{a}$ & Instrument & Reference\\
\hline
\endhead

\hline
\multicolumn{6}{r}{{Continued on next page}} \\
\endfoot

\hline
\endlastfoot
1350 -- 1480 & 1 & 0.56 & - & Rocket & \citet{Hayakawa1969}\\
2100 -- 2800 & $< 0.4$ & - & - & Rocket & \citet{Lillie1969}\\
1500 & 0.6 -- 1.0 & $< 0.5$ & - & OAO-2 & \citet{Witt1973}\\
2380 & 0.1 -- 0.3 & 0.75 & - & OAO-2 & \citet{Witt1973}\\
2200 & 0.3 -- 0.4 & 0.6 -- 0.9 & - & OAO-2 & \citet{Lillie1976}\\
1550 & 0.55 -- 0.65 & 0.6 -- 0.9 & - & OAO-2 & \citet{Lillie1976}\\
2740 & 0.68 & 0.5 & - & TD-1 & \citet{Morgan1978}\\
1530 & 0.5 & $0.95 <$ & - & Apollo 17 & \citet{Henry_dust1978}\\
1180 -- 1680 & - & - & 250 & Apollo 17$^{b}$ & \citet{Henry_ngp1978}\\
1230 -- 1680 & - & - & 253 -- 317 & Rocket$^{b}$ & \citet{Anderson1979}\\
1350 -- 1550 & - & - & 240 -- 360 & ASTP$^{b}$ & \citet{Paresce1979}\\
1350 -- 1550 & 0.5 & 0.5 & $<300$ & ASTP$^{b}$ & \citet{Paresce1980}\\
1565 & 0.5 & $>0.7$ & - & ASTP/TD-1 & \citet{Henry1981}\\
1200 -- 1670 & - & - & 100 -- 200 & Rocket$^{b}$ & \citet{Feldman_hotgas1981}\\
1690 & - & 0.6 -- 0.7 & 300 -- 690 & D2B$^{b}$ & \citet{Joubert1983}\\
2200 & - & - & 160 -- 360 & D2B$^{b}$ & \citet{Joubert1983}\\
1590 & - & - & $<550$ & Rocket$^{b}$ & \citet{Jakobsen1984}\\
1710 & - & - & $<900$ & Rocket$^{b}$ & \citet{Jakobsen1984}\\
2135 & - & - & $<1300$ & Rocket$^{b}$ & \citet{Jakobsen1984}\\
1700 -- 2850 & - & - & 200 -- 400 & Rocket & \citet{Tennyson1988}\\
1500 & $\geq 0.32$ & $\geq 0.5$ & 200 -- 300 & Rocket$^{b}$ & \citet{Onaka1991}\\
1415 -- 1835 & 0.13 -- 0.24 & $< 0.4$ & 110 & UVX & \citet{Hurwitz1991}\\
1500 & $>0.5$ & $>0.7$ & 200 -- 400 & UVX$^{b}$ & \citet{Henry1993}\\
1362 & 0.47 -- 0.70 & $<0.8$ & - & STS39 & \citet{Gordon1994}\\
1769 & 0.55 -- 0.72 & - & - & STS39 & \citet{Gordon1994}\\
1600 & 0.5 -- 0.7 & 0.35 -- 0.65 & - & UVX & \citet{Hurwitz1994}\\
1500 & 0.5 & 0.9 & 220 -- 380 & DE-1$^{b}$ & \citet{Witt1994}\\
1400 -- 1800 & 0.40 -- 0.50 & 0.58 -- 0.78 & 110 -- 210 & FAUST$^{b}$ & \citet{Witt1997}\\
1740 & 0.40 -- 0.50 & 0.67 -- 0.87 & 100 -- 300 & NUVIEWS$^{b}$ & \citet{Schiminovich2001}\\
1370 -- 1670 & 0.16 -- 0.56 & 0.30 -- 0.74 & - & Spear & \citet{Lee2008}\\
1565 & 0.4 & 0.7 & - & \galex\ & \citet{Sujatha2009}\\
1400 -- 1900 & 0.6 & 0.8 & 500 & DE-1$^{b}$ & \citet{Puthiyaveettil2010}\\
2365 & 0.37 -- 0.53 & 0.46 -- 0.66 & 40 -- 76 & \galex\ & \citet{Sujatha2010}\\
1565 & 0.23 -- 0.41 & 0.32 -- 0.70 & - & \galex\ & \citet{Sujatha2010}\\
1565 & - & 0.46 -- 0.70 & - & \galex\ & \citet{Murthy_halos2011}\\
2365 & - & 0.66 -- 0.78 & - & \galex\ & \citet{Murthy_halos2011}\\
1350 -- 1750 & 0.39 -- 0.45 & 0.25 -- 0.65 & - & SPEAR & \citet{Jo2012}\\
1360 -- 1680 & 0.37 -- 0.47 & 0.20 -- 0.58  & - & Spear & \citet{Lim2013}\\
1330 -- 1780 & 0.32 -- 0.44 & 0.40 -- 0.52 & - & Spear & \citet{Choi_spica2013}\\
1565 & 0.58 -- 0.66 & 0.73 -- 0.83 & 300 & \galex $^{b}$ & \citet{Hamden2013}\\
1565/2365 & 0.6 -- 0.7 & 0.2 -- 0.4 & - & \galex\ & \citet{Jyothy2015}\\
1350 -- 1700 & 0.35 & 0.6 & - & SPEAR & \citet{lim2015}\\
1565 & - & - & 286 -- 290 & \galex\ (NGP)$^{b}$& \citet{Akshaya2018}\\
2365 & - & - & 529 -- 533 & \galex\ (NGP)$^{b}$ & \citet{Akshaya2018}\\
1565 & - & - & 239 -- 243 & \galex\ (SGP)$^{b}$ & \citet{Akshaya2018}\\
2365 & - & - & 576 -- 582 & \galex\ (SGP)$^{b}$ & \citet{Akshaya2018}\\
1565 & 0.3 -- 0.5 & 0.7 -- 0.9 & 222 -- 258 & \galex $^{b}$& \citet{Akshaya2019}\\
2365 & 0.3 -- 0.5 & 0.4 -- 0.6 & 357 -- 431 & \galex $^{b}$ & \citet{Akshaya2019}\\
1565 & - & - & 240 -- 288 & Alice & \citet{Murthy2025_alice}\\
1565 & * & * & 260 -- 274 & \galex $^{b}$ & \cite{Murthy2025}\\
900 - 1600 & $a < 0.5$ & $g < 0.6$ & - & Alice & \citet{Murthy2026}\\
\hline
\multicolumn{6}{l}{a: \photu}\\
\multicolumn{6}{l}{b: Observation at high Galactic latitudes.}\\
\multicolumn{6}{l}{*: $a$ and $g$ formally ranged from 0 to 1 but were constrained to a narrow line.}\\
\end{longtable}

\end{appendix}

\clearpage
\bibliographystyle{abbrvnat}
\bibliography{murthy}

\end{document}